\documentstyle[sprocl]{article}

\input{epsf}
\def \bfgr #1{ \mbox {{\boldmath $#1$}}}

\def\PLB{{\em Phys. Lett.}  B}

\def\PRD{{\em Phys. Rev.} D}
\def\PRC{{\em Phys. Rev.} C}

\def\ZPA{{\em Z. Phys.} A}

\def\be{\begin{equation}}
\def\ee{\end{equation}}
\def\bea{\begin{eqnarray}}
\def\eea{\end{eqnarray}}

\begin{document}

\title{NUCLEON AND NUCLEAR SPIN STRUCTURE FUNCTIONS
\footnote{Invited talk at the Conference ``Perspectives in Hadronic
Physics'', Trieste, May 12-16, 1997. Presented by S. Scopetta}}

\author{S. SCOPETTA}

\address{
Departament de Fisica Teorica, 
 Universitat de Val\`encia, E-46100 Burjassot (Val\`encia), Spain}

\author{A.Yu. UMNIKOV}

\address{
INFN, Sezione di Perugia, via A. Pascoli, I-06100 Perugia, Italy.}

\author{C. CIOFI DEGLI ATTI}

\address{Department of Physics, University of Perugia, and
INFN, Sezione di Perugia, I-06100 Perugia, Italy.}

\author{L.P. KAPTARI}

\address{
   Bogoliubov Lab.  of Theor.  Phys., Joint Institute for Nuclear Research,
   Dubna, Russia.}


\maketitle\abstracts{
Nuclear effects in polarized inelastic electron scattering
off polarized $^3He$ and polarized $^2H$ are discussed;
in the resonance region, 
Fermi motion effects are found to be much larger than
in deep inelastic scattering. 
It is shown that improperly describing nuclear dynamics
would lead to the extraction of unreliable
neutron spin structure functions; 
on the other hand side,
simple and workable
equations, relating the Gerasimov -- Drell -- Hearn Integral 
for the neutron
to the corresponding quantity for $^3He$ and $^2H$, are proposed.
Nuclear effects in the recent E143 data  
are estimated by a proper procedure. }

\section{Introduction}

The measurement of the polarized nucleon Spin Structure Functions 
(SSF) 
$g_1$ and $g_2$ 
in the resonance 
region allows one to check the helicity structure 
of the photon -- nucleon
coupling between the Deep Inelastic Scattering (DIS) region 
and the real photon limit~\cite{drech}.
Recently it has been proposed at TJNAF to
study the SSF in a wide range of 
energy ($0.2~GeV \leq \nu \leq 3~GeV$) and momentum
($0.15~GeV^2\leq Q^2 \leq 2~GeV^2$) transfers,
for both the proton\cite{cebp} and the neutron,
using in the latter case
polarized $^2H$  and $^3He$ 
targets\cite{cebd,ceb3}.

For a  hadronic target $A$,
the SSF $g_1^A(\nu,Q^2)$ and $g_2^A(\nu,Q^2)$ are 
experimentally obtained by measuring 
longitudinal and transverse asymmetries\cite{IOFFE}.
A relevant quantity related to the SSF's $g_{1(2)}$ is the following integral
\begin{eqnarray}\label{gdh}
I^A(Q^2) =
{8 \pi^2 \alpha \over m}
\int\limits_{\nu_{th}}^\infty
{ d \nu \over \nu} 
{ \left ( 1 + {Q^2 \over \nu^2} \right ) \over K} 
{\cal G}_1^A(\nu,Q^2)~,
\end{eqnarray}
where $\nu_{th}=(Q^2+2m_{\pi}m+m_{\pi}^2)/2m$ is the threshold 
energy for the pion-e\-le\-ctro\-pro\-du\-ction off
the the nucleon, $m$ is the nucleon mass, $m_{\pi}$ is the pion mass, 
$K$ is the photon flux,  $\alpha$  is the fine structure constant and 
${{\cal G}}_1^A(\nu,Q^2)$ reads as:
\begin{eqnarray}\label{g1s}
{{\cal G}}_1^A(\nu,Q^2) & = & 
{ 1 \over \left( 1 + {Q^2 \over \nu^2} \right)} 
\left( g_1^A(\nu,Q^2) - {Q^2 \over \nu^2} g_2^A(\nu,Q^2) \right)~.
\end{eqnarray}
Two  features of ${{\cal G}}_1^A(\nu,Q^2)$
have to be mentioned:
(i)  when
multiplied by the quantity ${-\frac {2 \nu}{M_A}}(1+\frac {Q^2}{\nu^2})$
, $M_A$ being the target mass, it
coincides with the usual transverse response, $R_{T'}$ (see e.g.\cite {prc}); and
(ii) in the DIS limit ($Q^2 \rightarrow \infty$, $\nu \rightarrow \infty$, 
$Q^2/\nu$ fixed),
${\cal G}_1^A$  coincides with the SSF, $g_1^A(\nu,Q^2)$.

For any spin ${1 \over 2}$ hadronic target, the SSF $g_1^A$
and $g_2^A$ 
read as follows\cite{IOFFE}
\begin{equation}\label{g1}
g_1^A(\nu,Q^2)=\frac {M_A K}{8\pi^2 \alpha(1+\frac {Q^2}{\nu^2})} \left [
\Delta\sigma^A(\nu,Q^2)
+\frac {2\sqrt {Q^2}}{\nu}
\sigma_{TL}^A(\nu,Q^2)\right ]~,
\end{equation}
\begin{equation}\label{g2}
g_2^A(\nu,Q^2)=\frac {M_A K}{8\pi^2\alpha(1+\frac {Q^2}{\nu^2})}\left [
\frac {2\nu}{\sqrt{Q^2}}\sigma_{TL}^A(\nu,Q^2)-
\Delta\sigma^A(\nu,Q^2)
\right ]~,
\end{equation}
where $ \Delta\sigma^A(\nu,Q^2) = 
\sigma_{1/2}^A(\nu,Q^2)-\sigma_{3/2}^A(\nu,Q^2) $,
 $\sigma_{1/2(3/2)}^A(\nu,Q^2)$ is the cross section
for photon -- hadron scattering with 
total helicity ${1/2\,(3/2)}$,
$\sigma_{TL}^A(\nu,Q^2)$ is the transverse -- longitudinal
interference cross section.  Thus Eq.~(\ref{g1s}) becomes: 
\begin{eqnarray}\label{g1nn}
{{\cal G}}_1^A(\nu,Q^2)
& = & \frac {M_A K}{8\pi^2 \alpha(1+\frac {Q^2}{\nu^2})} \left [
\sigma_{1/2}^A(\nu,Q^2)-\sigma_{3/2}^A(\nu,Q^2) \right ]~.
\end{eqnarray} 

For a nucleon target ($A=N$, $N=n$ or $p$),
the integral $I^A$ coincides with
the Gerasimov -- Drell -- Hearn (GDH)
integral, $I^N_{GDH}(Q^2)$, which, in the real photon limit,
gives the GDH
Sum Rule\cite{DHG66}:
\begin{eqnarray}\label{sr}
I^N_{GDH}(Q^2=0) & = &
\int\limits^{\infty}_{\nu_{th}}
\frac {d\nu}{\nu}
\left ( \sigma^N_{1/2}(\nu,Q^2=0)-\sigma^N_{3/2}(\nu,Q^2=0) \right ) 
= -\frac {2 \pi ^2 \alpha }{m^2}\kappa_N^2
\cr 
& \simeq & 
\cases{
-0.53 \quad GeV^{-2} & 
for protons\cr
-0.60 \quad GeV^{-2} & for neutrons \cr}
\end{eqnarray}
where $\kappa_N$ is the anomalous magnetic moment of the nucleon.

An important observation \cite{ans} 
can be made about the $Q^2$ evolution
of $I_{GDH}^N$.  Since 
in the large $Q^2$ limit ${\cal G}_1^N$
coincides with  $g_1^N$, one can evaluate (\ref{gdh}) 
for the proton, using the results\cite{EMC}
at $Q^2 \simeq 10 ~GeV^2$, which gives  
$I_{GDH}^p(Q^2)
\simeq 0.14/Q^2$. This
result, if compared with (\ref{sr}),
provides evidence of a sharp modification
in the helicity structure of the $\gamma p$ coupling between the real
photon limit (\ref{sr}) and the DIS region, which lead
to a change of sign of $I_{GDH}^p(Q^2)$ at some value of $Q^2$.
 In order to understand this behavior,
much theoretical work has been produced\cite{ik73,burk},
both in the real photon limit  and at  
finite values of $Q^2$. 
The electroexcitation 
of nucleon resonances is evidently the main reason for this
evolution of the integral, $I_{GDH}^p(Q^2)$. 
Therefore,   experimental investigation of the
low-$Q^2$ evolution 
of $I^N_{GDH}$ is   of great relevance, in particular
for the neutron, for which 
several analyses of the available unpolarized data for
photoproduction and low-$Q^2$ electroproduction
disagree with the prediction (\ref{sr}), 
while similar estimates seem to give the correct 
value for the proton.

Since ${\cal G}_1^n(\nu,Q^2)$ will be extracted 
from experimental data on
$\vec{^2H}$   and $\vec{^3He}$,
corrections due to nuclear effects have to be introduced.
The goal of this talk is to illustrate 
the relevance of
these corrections 
in the resonance region.
Our results for  
$\vec{^3He}$  and $\vec{^2H}$
 are presented in the following two sections\cite{plbmio,plb}.
The recent deuteron data
of the E143 collaboration\cite{e143} are used as an example
to test
the method
proposed in\cite{kanna}
for the extraction of the neutron SSF from the nuclear ones. 

\section{The process $\vec{^3He}(\vec e,e')X$ in the resonance region}

Using the
convolution model for the  nuclear SSF $g_{1(2)}^A$
described in\cite{prc,pg,sa},
we obtain\cite{plbmio}
\begin{eqnarray}
{\cal G}^A_1 \left(\nu, Q^2 \right ) & = & 
 \sum\nolimits\limits_{N=p,n} \int \nolimits \nolimits~dz
\int \nolimits \nolimits~dE \int \nolimits
\nolimits~d{\vec p}~ {m \over E_p}~{m \nu  \over p \cdot q}~
\left \{  {\cal G}_{1}^{N}\!\left(\nu',{Q}^{2} \right) 
~{\hbox{\Large\it P}}_{\parallel}^{~N}\!\left( {\vec p},E\right)
\right. 
\cr
& & 
\left. 
+~{\hbox{\Large\it T}}^{~N}\!\left( {\vec p},E,Q^2\right) \right \} 
\delta \left(z+{m^2-p \cdot p\over 2m \nu} -{q \cdot p \over
m\nu}\right) ~,
 \label{eq11}
\end{eqnarray}

\noindent where $E$ is the removal energy
of a nucleon with momentum $\vec p$,
$p\equiv(M_A-\sqrt{(E+M_{A}-m)^2+|\vec
p|^2},~\vec p)$ is
the nucleon 4-momentum,
$q$ is the 4-momentum transfer, 
$~E_p=\sqrt{m^2+|\vec{p}|^2}$,
${\hbox{\Large\it P}}_{\parallel}^{~N}\!\left(  \vec {p}, E\right)$ 
and ${\hbox{\Large\it T}}^{~N}\!\left(\vec {p},E, Q^2\right)$
are related to the elements of the 2x2 matrix,
which represents the spin dependent spectral function\cite{prc}.
\begin{figure}
\vskip 3.0cm
\includegraphics{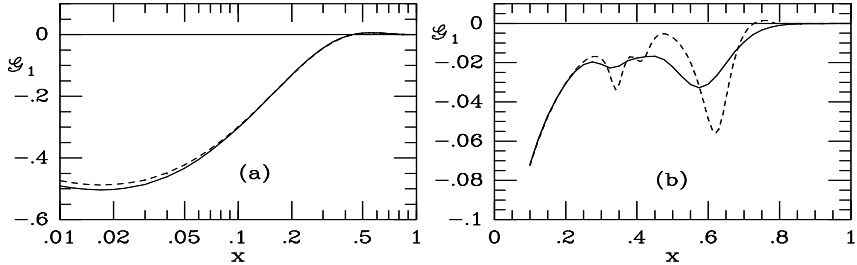}
\caption{
${\cal G}_1^{^3He}$ 
in DIS ($Q^2=\,10\,GeV^2$) (a)
\protect\cite{pg}, and in resonance ($Q^2=1~GeV^2$) (b) regions,
obtained by considering Fermi
motion and binding (full). The dashed curve represents the same functions
obtained considering the proton and neutron effective polarization in $^3He$
as the only relevant nuclear effects (Eq. (\protect \ref{gmod2})).
\label{fig:one}}
\end{figure}

The elements of this matrix are
\begin{eqnarray}
P_{\sigma,
\sigma',\cal{M}}^{N} ({\vec{p},E}) & = & \sum\nolimits\limits_{{f}_{A-1}}
{}~\langle{\vec{p},\sigma;\psi
}_{A-1}^{f} |{\psi }_{J\cal{M}}\rangle~ \langle{\psi
}_{J\cal{M}}|{\psi }_{A-1}^{f};\vec{p},\sigma '\rangle ~
\cr
& \times & \delta (E-{E}_{A-1}^{f}+{E}_{A}) 
\label{eq9}
\medskip
\end{eqnarray}
where $|{\psi}_{J\cal{M}}\rangle$ is the ground state of the polarized target 
nucleus, $|{\psi }_{A-1}^{f}\rangle$ is the eigenstate of the (A-1) nucleon
system with eigenvalue
 $E_{A-1}^f$, and  $|\vec{p},\sigma\rangle$ 
is the plane wave describing the struck nucleon in  continuum.
The relevant quantity is ${\hbox{\Large\it
P}}_{\parallel}^{~N}\!\left(\vec p,E\right)$=
$P^N_{ {1 \over 2} {1 \over 2}, {1 \over 2} }( \vec p, E)
- P^N_{ - {1 \over 2} - {1 \over 2}, {1 \over 2} }( \vec p, E)$,
i. e. the difference between
the spectral functions with nucleon spin
parallel or antiparallel 
to the nucleus spin. It provides
 the effective nucleon polarizations $p_N$ 
produced by the $S'$ and $D$ waves in the ground state of $^3He$:
\begin{eqnarray} 
p_{N}& = & \int dE~\int d\vec p ~{\hbox{\Large\it
P}}_{\parallel}^{~N}\!\left(\vec p,E\right)~.
\label{pol}
\medskip
\end{eqnarray}
The spectral function of \cite{prc} gives $p_{p(n)}=-0.030$ $(0.88)$,
in agreement with world calculations on the three body systems\cite{friar},
$p_{p(n)}=-0.028 \pm 0.004$ $(0.86 \pm 0.02)$.
The term
${\hbox{\Large\it T}}^{~N}\!\left(\vec {p} ,
E, Q^2\right)$
depends also upon  
$P^N_{ {1 \over 2} -{1 \over 2}, {1 \over 2} }( \vec p, E)$, 
as well as upon a proper combination of the SSF's
$g_1$ and $g_2$\cite{pg}. However, it is
of the order $|\vec {p}/m|$, and thus
gives a very small contribution
both in the DIS\cite{pg} as well as in the present
calculation of the resonance region.
For this reason it will be omitted hereafter. 
The quantity
${\cal G}_1^N$ 
which appears in
${\cal G}_1^{^3He}$ (Eq. (\ref{eq11}))
is defined in terms
of the polarized electroproduction cross-sections 
$\sigma_{1/2(3/2)}^N$ (cf. Eq. (\ref{g1nn})).
Since the first few data for ${\cal G}_1^N$
appeared very recently \cite{e143}, 
we use a theoretical model \cite{burk}, where the
contributions of the resonances 
$P_{33}(1232)$, $D_{13}(1520)$,
$S_{11}(1535)$ and $F_{15}(1680)$ have been parametrized  using 
the existing experimental data for unpolarized
electroproduction. Other 
resonant states have been added  using the
predictions of a single quark transition model, and 
the single pion Born term background has also been included.
Using the above models for
${\hbox{\Large\it
P}}_{\parallel}^{~N}\!\left(\vec p,E\right)$ and
${\cal G}_1^N$,
we have  calculated
${\cal G}_1^{^3He}$ in the resonance region. The results are presented in 
Fig. 1,
where they are compared
with the results for the DIS limit \cite{pg}. 
In the latter case 
${\cal G}_1^A(\nu,Q^2)= g_1^A(\nu,Q^2)=\frac {M_A K}{8\pi^2 \alpha}
\left [\sigma_{1/2}^A(\nu,Q^2)-\sigma_{3/2}^A(\nu,Q^2) \right ]$~,
and Eq. (\ref{eq11}) reduces to the well-known convolution formula:
\begin{eqnarray}
g_1^A(x,Q^2) &  = &  \sum_N \int \limits_x ^{M_A/m} dz
~{1 \over z}~ g_1^N \left( {x \over z}, Q^2 \right)
~G^N(z)~~, \label{fin}
\medskip
\end{eqnarray}
where $G^N(z)$ is the light-cone momentum distribution
\begin{eqnarray}
G^N(z) & = &  \int dE\, \int d {\vec p}~{\hbox{\Large\it
P}}_{\parallel}^{~N}\!\left(\vec p,E\right)~
\delta \left(z - {p^+ \over M} \right)~~~ \label{lux}
\medskip
\end{eqnarray}
with  $p^+=(p^0 \nu - \vec p \cdot \vec q )/ |\vec q|$ 
being  the light-cone
momentum component.
Fig.~1 shows that in the DIS case
the following equation
\begin{equation}
{\cal G}_1^{^3He}(x,Q^2)  \approx  2p_p {\cal G}_1^p (x,Q^2) + p_n 
{\cal G}_1^n(x,Q^2)~,
\label{gmod2}
\medskip
\end{equation}
approximates very well the convolution formula,
at least for $x \leq~0.8$ \cite{pg}. 
The same does not hold in 
the resonance region,
where nuclear effects 
turn out to be important,
with
the Fermi motion significantly
broadening and damping the peaks associated
with the excitation of
the prominent resonant states. 
Therefore, 
Eq. (\ref{gmod2}) can be considered as a workable formula for 
extracting the DIS $g_1^n(x,Q^2)$ from experimental data on
$g_1^{^3He}(x,Q^2)$, but
in the resonance region it appears to be of little help.
We stress that the importance of nuclear effects in the
resonance region is a well-known feature of the unpolarized
scattering as well.
Nevertheless, we have found that the proton contribution to the nuclear
${\cal G}_1^{^3He}(x,Q^2)$ is not large, and therefore $^3He$ is
a good effective polarized neutron target also in 
the resonance region.

\begin{figure}
\vskip 4.5cm
\includegraphics{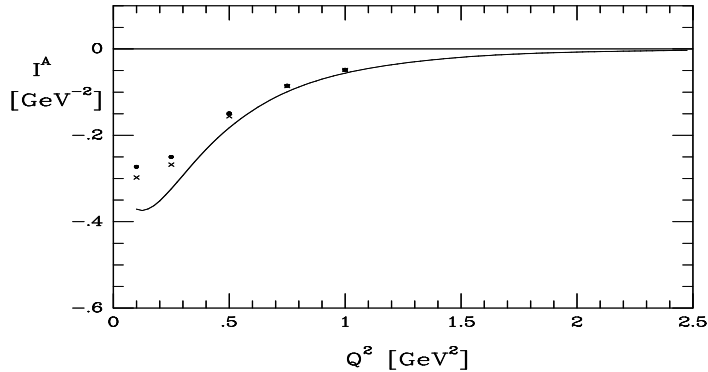}
\caption{
The integral ${I}^{^3{He}}(Q^2)$,
Eq. (5) (dots),
compared 
with the approximation (13) (crosses) and with
$I^n(Q^2)$ \protect\cite{burk} (solid line).
}
\end{figure}

Let us now
discuss the role of nuclear effects on the 
integral Eq. (\ref{gdh}). In Fig. 2
the  integral, calculated placing 
Eq. (\ref{eq11}) in Eq. (\ref{gdh}), is compared with the
following expression
\begin{eqnarray} 
\label{itiln3}
{\tilde I}^{^3He}(Q^2) 
& = &
2p_p I^p(Q^2) 
+{p_n}  I^{n}(Q^2)~,
\medskip 
\end{eqnarray}
which represents the integral (\ref{gdh}) within
the assumption
that Fermi motion and binding can be 
disregarded and that nuclear effects in polarized $^3He$ are
 only due to 
the effective nucleon polarizations. It can be seen 
that Eq. (13) approximates Eq. (\ref{gdh}) with an accuracy better than
 5\%. From the same figure it can also be seen that,
because of the effective nucleon polarizations, the integral (\ref{gdh})
for the free neutron predicted by \cite{burk}, 
noticibly differs
from  $I^{^3He}(Q^2)$. Thus the 
nuclear structure effects are very large in the nuclear SSF
${\cal G}_1^A(\nu,Q^2)$ (cf. Fig. 1 (b)), but 
in the integral $I^{^3He}(Q^2)$ these effects can be 
approximately accounted for
 by the effective nucleon polarizations
(cf. Fig. 2).
This result is understood \cite{plbmio} by 
performing a series expansion 
of the r.h.s. of 
Eq. (\ref{eq11})
in the variable $z$ 
around the non relativistic value $z=1$. 

To sum up,
we have shown that in the resonance region
${\cal G}_1^{^3He}(x,Q^2) \neq p_n {\cal G}_1^n(x,Q^2)$ $+2p_p
{\cal G}_1^p(x,Q^2)$, but the integrals of the two quantities
are very similar.

\section{The process $\vec{^2H}(\vec e,e')X$ in the resonance region}

 The nucleon contribution to the deuteron structure functions
 is usually calculated by weighting the amplitude
 of  electron scattering on the nucleon with the wave function of nucleon
 in the deuteron (for recent developments see
 e.g.~\cite{ourphyslet,mt2}
 and references therein).
 For the SSF the most important effects 
 are the Fermi motion  and the depolarizing effect of
 the D-wave.
 Additional effects, such as off-mass-shell effects 
 or nucleon deformation,
 are found to be small~\cite{mst}.
 For finite values of $Q^2$ and $\nu$, 
Eq. (\ref{g1s}) for the 
 deuteron 
reads as follows\cite{plb}
  \begin{eqnarray} 
{\cal G}_1^D(x,Q^2) && = 
\int \frac{d^3{\bfgr p}}{(2\pi)^3}
\frac{m\nu}{pq}
 {\cal G}_1^N\left (x^*,Q^2 \right )
\left ( 1+\frac{\xi (x,Q^2) p_3}{m}\right )
\cr
&& 
\times  
\left ( \Psi_D^{M+}({\bfgr p}) S_z \Psi_D^M({\bfgr p}) \right)_{M=1}
\label{gd}
\cr 
&& =  
\int\limits_{z_{min}(x,Q^2)}^{z_{max}(x,Q^2)}
\frac{dz}{z} {\cal G}_1^N(x/z,Q^2)
 \vec f_D(z,\xi(x,Q^2)),
 \label{conv} 
\end{eqnarray}
where ${\cal G}_1^N=({\cal G}_1^p+ {\cal G}_1^n)/2$ is the isoscalar nucleon 
response 
given by Eq. (\ref{g1nn})
and $\Psi_D^M({\bfgr p})$\ the deuteron wave function  with spin
 projection $M$. In the rest-frame
of the deuteron, with ${\bfgr q}$ opposite the z-axis,
the kinematical variables are defined as follows:
\begin{eqnarray}
&& pq = \nu (p_0 + \xi(x,Q^2)p_3),\quad p_0 = m+\epsilon_D- {\bfgr p}^2/2m,\\
&&\!\!\!\!\!\!\!\!\!\!\!\!\!\!
\xi \equiv q_3/\nu= |{\bfgr q}|/\nu = \sqrt{1+4m^2x^2/Q^2},
\quad Q^2 \equiv -q^2,\quad x^* = Q^2/2pq,
\end{eqnarray}
where $\epsilon_D=-2.2246~MeV$ is the deuteron binding energy.

The limits ${z_{min (max)}(x,Q^2)}$ are defined
to provide an integration over the physical region of
momentum in (\ref{gd}) and to take into account the 
pion production
threshold  in the virtual photon-virtual nucleon
scattering.
Eq. (\ref{conv}) has the correct limit in
DIS.
In this case: $\xi(x,Q^2) \to 1, \quad z_{min} 
\to x, \quad z_{max}\to M_D/m$, and the usual convolution
formula for the deuteron SSF $g_1^D(x,Q^2)$
is recovered ~\cite{ourphyslet,mt2}: 
 \begin{eqnarray} 
g_1^D(x,Q^2)=\int\limits_{x}^{M_D/m}\frac{dz}{z} g_1^N(x/z,Q^2)
 \vec f_D(z).
 \label{convdis} 
\end{eqnarray}
(cf. Eq. (12) for the $^3He$ case).
Equation (\ref{convdis}) defines the spin-dependent  
``effective distribution
of the nucleons", $\vec f_D$, which describes the bulk of nuclear 
effects in  $g_1^D$. The main features of the distribution function, 
$\vec f_D(z)$, are a sharp maximum at $z = 1+\epsilon_D/2m\approx 
0.999$ and a normalization given by $(1-3/2P_D)$
 ($P_D$ being the weight of the D-wave in the deuteron). 
As a result in the region of medium values of $x\sim 0.2-0.6$ 
the deuteron SF $g_1^D(x)$ is slightly
 suppressed  by Fermi motion and binding effects, 
compared to $(1-3/2P_D)\times g_1^N(x)$. However,
the magnitude of this suppression is small ($\sim 1\%$) and this is
why it is phenomenologically acceptable to extract the neutron SF
from the deuteron and proton data by making use of the following
approximate formula:
 \begin{equation} g_1^D(x,Q^2)\approx \left
 (1-\frac{3}{2}P_D\right )( g_1^n(x,Q^2) + g_1^p(x,Q^2))/2~, 
 \label{extract}
 \end{equation}
(cf. Eq. (14) for the $^3He$ case).
In addition, when integrated over $x$, eqs. (\ref{convdis}) and
(\ref{extract}) give exactly the same result ($\Gamma = \int dx g_1(x)$), i.e.
 \begin{equation}
  \Gamma_D(Q^2) = \left(1-\frac{3}{2}P_D\right )
(\Gamma_n(Q^2) + \Gamma_p(Q^2))/2,
 \label{gamma}
 \end{equation}
which allows one to obtain {\em exactly} the integral of the neutron SF
$\Gamma_n$ knowing the deuteron and proton integrals,
without solving (\ref{convdis}).

As in the $^3He$ case, the equations
at finite values of $Q^2$ and $\nu$ are more sophisticated than 
the corresponding equations in the deep inelastic limit.
 In particular, Eq. (\ref{gd}) does not represent
 a ``convolution formula" in the 
 usual sense, since the effective distribution function  $\vec f_D$
 and the integration limits
 are also functions of $x$. This circumstance immediately leads to the
 conclusion that, in principle, when integrals
 of the SF are considered, the effective distribution
 can not be integrated out to get a factor similar to $(1-3/2P_D)$
in (\ref{gamma}).
 Another interesting feature of Eq. (\ref{gd})
is the $Q^2$-dependence of  $\vec f_D$ and 
${z_{min,(max)}(x,Q^2)}$. If we again limit ourselves to
the discussion of the integrals of SF, one concludes
that the $Q^2$-dependence
of such an integral is governed by both the QCD-evolution of the nucleon
SF and the kinematical $Q^2$-dependence of the effective distribution
of nucleons. 
Thus, in principle, 
in the non-asymptotic regime, equation
(\ref{gamma}), does not hold. 

We have performed a realistic calculation of Eq. (\ref{gd}). 
As in the $^3He$ case, in our numerical estimates we have
evaluated ${\cal G}_1^N$ in (\ref{gd}) using the elementary
cross sections 
for the proton and neutron given in~\cite{burk}.
Using the Bonn potential model for
the deuteron wave function~\cite{bonn}, we have carried out 
a realistic calculation of
${\cal G}_1^D(x,Q^2)$, (\ref{gd}), in the region of nucleon resonances. 
\begin{figure}
\vskip 4.5cm
\includegraphics{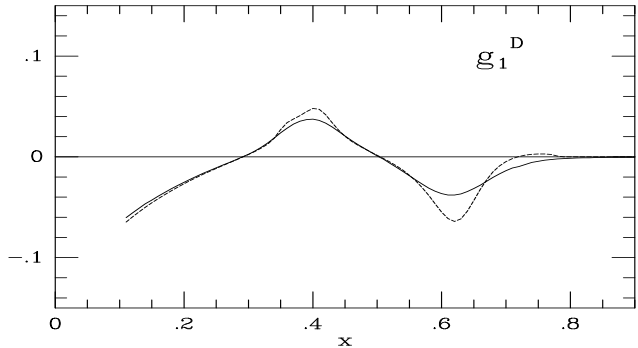}
\caption{
${\cal G}_1(x,Q^2)$ at $Q^2$=1 GeV$^2$ for the deuteron SF (solid line),
compared with the corresponding quantity for the isoscalar nucleon 
(dotted line), used as input in
the calculation of eq.~(\protect\ref{conv}). }
\end{figure}
  In Fig. 3, the obtained 
${\cal G}_1^D(x,Q^2)$ 
at $Q^2 = 1.0$ GeV$^2$, 
is compared with the input of the calculation, i.e. the isoscalar
 nucleon response, ${\cal G}_1^N(x,Q^2)$.
It can be seen that
the role of nuclear effects in the 
resonance region is much larger (up to $\sim 50\%$ in
the maxima of the resonances),
than in the deep inelastic regime ($\sim 7-9\%$, depending upon
the models~\cite{ourphyslet,mt2}, with resulting
 $\sim 6-7 \%$ from the
 depolarization factor  $(1-3/2P_D)$ and  $\sim 1-2\%$ from the
 binding effects and Fermi motion).
As in the $^3He$ case, such a drastic effect is a consequence of 
the presence of the narrow resonance peaks in the nucleon 
${\cal G}_1^N(x,Q^2)$.


 Let us now discuss
 the role of nuclear corrections in the 
 analysis of the integrals of the SSF, such as the GDH Integral.

 An important observation has been made
 in DIS, namely 
 the {\em exact} formula (\ref{convdis}) and the {\em approximate} formula
 (\ref{extract})  give the same result for the integral of the neutron
  structure function, $g_1^n(x,Q^2 \gg m^2)$ (see eq.~(\ref{gamma})).
 The applicability of the approximate formula in DIS
is based on the conservation
of the norm of the distribution $\vec f(z)$  by the convolution formula
(\ref{convdis}).
This circumstance can not be immediately
extended to the case of the resonance region, since: (i)
the covolution is broken in eq. (\ref{conv}) and (ii) 
the normalization of the function $\vec f(z,x,Q^2)$ is different from
the one of $\vec f(z)$.  The integral of the distribution $\vec f(z,x,Q^2)$
 represents the
``effective number" of nucleons ``seen" 
by the virtual photon in the process when
the virtual photon is absorbed by the nucleon
and at least one pion is produced
in the final state (it   is less than 1 at low $Q^2$
and $x \to x_{max}$).

However, 
the use of the formula~(\ref{gamma}) in the
resonance region gives results numerically very 
close to the integration of the exact equation~(\ref{conv}).
As explained in\cite{plb},  
this is a consequence of the smallness of the effects breaking 
the convolution in eq.~(\ref{conv}).

We have found that the integrals of the SF,
 such as the GDH  Sum Rule,
 can be estimated with accuracy better
 than 3\% by the simple formula (\ref{gamma}) which is also valid in 
 deep inelastic region.

\section{Neutron SSF from nuclear data}

 At this point, we have observed
 that nuclear effects in the resonance region are
 very specific and therefore approximate formulae,
 such as (\ref{gmod2}) for $^3He$ or (\ref{extract})
 for $^2H$,
 do not work even for crude extraction
 of the neutron response. 
  Obviously, another 
 method of extracting should be used. 

In ref.~\cite{kanna} a rigorous
 procedure of solving
 eq. (\ref{convdis}) for the unknown neutron SSF has been proposed and
 applied in the deep inelastic region. 
 It has been shown that this method,
which works for both 
spin-independent and
spin-dependent SSF, 
in principle allows one to
extract the neutron SSF exactly, requiring
only the analyticity of
SSF. It can also be applied by a minor modification
to the extraction of the SSF at finite $Q^2$. 
Such a procedure can be applied both to $^2H$ and $^3He$ targets;
here only the deuteron case will be discussed.

  The basic idea is to replace the integral equation
 (\ref{conv}) by a set of linear algebraic equations, $KG_N =
 G_D$, where $K$ is a square  matrix
(depending upon the deuteron model),
 $G_D$ is a vector containing the available 
 ${\cal G}_1^D$ data and $G_N$ is 
a vector of unknown solutions. 
 Changing the integration variable
 in (\ref{conv}), $\tau =
 x/z$, we get
 \begin{eqnarray} 
{\cal G}_1^D(x,Q^2)=\int\limits_{\tau_{min}(x,Q^2)}^{\tau_{max}(x,Q^2)}
d\tau {\cal G}_1^N(\tau,Q^2)
\frac{1}{\tau} \vec f_D(x/\tau,\xi(x,Q^2)),
 \label{conv2} 
\end{eqnarray}
where $\tau_{min}(x,Q^2)= x/z_{max}(x,Q^2)$,
$\tau_{max}(x,Q^2)=x_{max}(Q^2)/z_{min}(x,Q^2)$ and $x_{max}(Q^2)$ is 
defined by the pion production threshold in virtual photon-nucleon
scattering. Let us assume that ${\cal G}_1^D$ has been measured
experimentally in the interval $(x_1,x_2)$ and that a reasonable 
parametrization for it can be obtained in this interval. 
 Then, dividing both intervals $(x_1,x_2)$ and $(\tau_{min}, 
\tau_{max})$
 into $N$ small parts, one may write:
  \begin{eqnarray} && {\cal G}_1^D(x_i,Q^2)\approx
 \sum_{j=1}^N {\cal G}_1^N(\tilde\tau_j,Q^2)\int\limits_{\tau_j}^{\tau_{j+1}}
\frac{1}{\tau} \vec f_D(x_i/\tau,Q^2) d\tau, \quad i=1\ldots N,
\label{mat}
\end{eqnarray}
where $\tilde\tau_j = \tau_{min}+h(j-1/2)$ and 
 $h=(\tau_{max}-\tau_{min})/N$. Equation (\ref{mat})
 is already explicitly of the form 
 $G_D=K G_N$, therefore usual linear algebra methods can be
applied to solve it.
 
Note that the range of variation of
$\tau$ is larger than the one for $x$.
Therefore the experimental knowledge of 
${\cal G}_1^D$ in the interval $(x_1,x_2)$ provides
information about the neutron in a wider interval (for example, in
deep inelastic regime $\tau_{min}\approx x/2$ and
$\tau_{max}=1$). However, extracting information beyond the interval
 $\tilde \tau_{min}=x_1$ to
$\tilde\tau_{max}=x_2$ is almost impossible in view of the structure
of the kernel of eq.~(\ref{conv2}) and
the kinematical condition of 
planned experimental data~\cite{kanna}. We have to redefine 
the kernel  of eq.~(\ref{conv2}) to incorporate new limits
of integration  $\tilde \tau_{min}=x_1$ and
$\tilde\tau_{max}=x_2$~\cite{kanna}. The experimental errorbars
of the nuclear data 
can be related to errors in
the extracted structure function of the nucleon.

\begin{figure}
\vskip 6.cm
\includegraphics{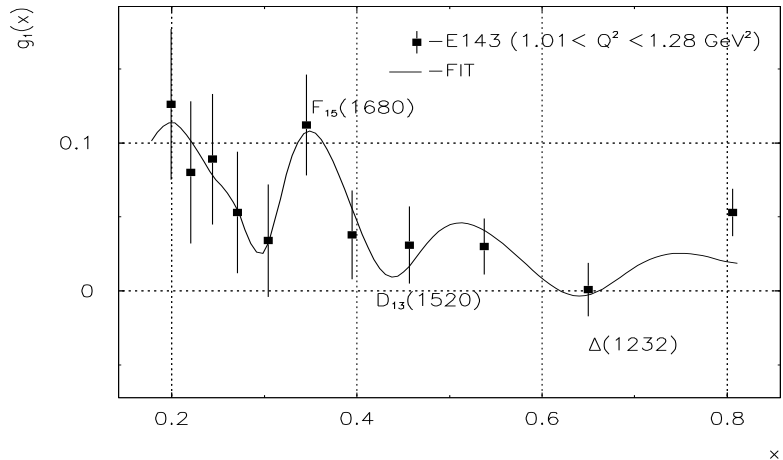}
\caption{
The analytical fit of the deuteron E143 data \protect\cite{e143}
used as input in our example of extraction }
\end{figure} 

To show how this procedure works in practice, we have obtained
${\cal G}_1^N$ of the isoscalar nucleon solving Eq. (\ref{gd}),
using in the left hand side an analytical fit of one set
of the recent E143 deuteron data\cite{e143} (corresponding to
1.01 $\leq Q^2 \leq$  1.28 GeV$^2$, see Fig. 4). 
It should be noticed that in the experimental
analysis the difference between $g_1$ and ${\cal G}_1$ 
(cf. Eq. (\ref{g1s})) has been neglected, and therefore
the data refer to the SSF $g_1^D$ of the deuteron.
\begin{figure}
\vskip 6.cm
\includegraphics{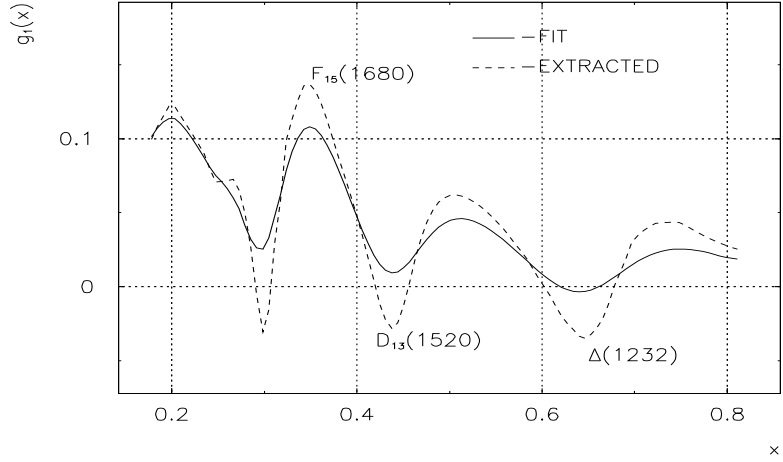}
\caption{
The extracted isoscalar nucleon SSF (dashed line) compared
with our fit of the E143 data \protect\cite{e143} (full line), used
as input in our calculation.}
\end{figure}
\begin{figure}
\vskip 6.cm
\includegraphics{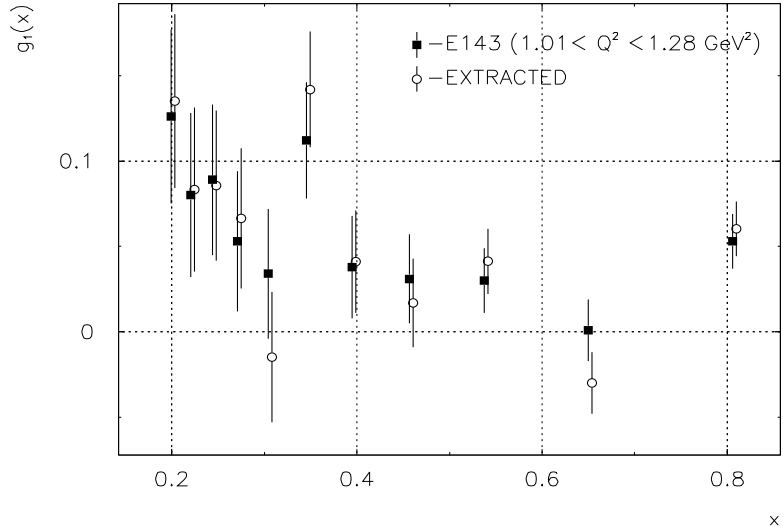}
\caption{
The extracted ``data'' of the isoscalar nucleon SSF (open circles)
compared with the experimental deuteron ones (filled squares).
The extracted points have been slightly shifted, in order to
distinguish their errorbars from the the experimental ones.}
\end{figure}
Resonance structures
corresponding to the three main resonances
have been included in obtaining the fit. However, no further physical
assumptions have been made, and the behavior
of the fitting curve is basically forced by the data,
whose errorbars are large. Therefore, our analysis is 
only a crude estimate of the possible nuclear effects,
and presently it should be considered as an example of how 
the extraction 
method could be applied. 
Once we have assumed that our fit describes the nuclear data
fairly well, Eq. (\ref{gd}) can be solved using the
method described in this section.
The results of the calculation are shown in Figs. 5 and 6.
It can be seen that nuclear effects are large, and that the
extraction procedure allows one to discover features of the 
isoscalar nucleon SSF which, due to nuclear effects, 
are not apparent in the initial nuclear SSF. 
The neutron SSF could be obtained by subtracting
the proton data from the isoscalar SSF, and provided the
errorbars on ${\cal G}_1^D$ are reduced, 
this method will provide us with 
a reliable neutron SSF.
  
\section{Conclusions}

We have shown that the effects of nuclear
structure  in the extraction of the neutron SSF in the
resonance region are much more important than in DIS.
We have explained how the correct neutron SSF can be firmly 
extracted from the combined deuteron and proton data.
As for the integrals of the SSF,
the estimates are easier, and they can be carried out by
using the simple procedure similar to the DIS case.


\section*{Acknowledgments}

Two of us (S.S. and A.U.) thanks the Organizers of the Conference for the invitation,
and 
S.S. thanks the TMR programme of the European Commission ERB FMRX-CT96-008
for partial financial support.

\end{document}